\documentclass[aip,jcp,reprint,superscriptaddress]{revtex4-2}%
\usepackage{graphicx}
\usepackage{xcolor}
\usepackage{tabularx}
\usepackage{dcolumn}
\usepackage{longtable}
\usepackage{tensor}
\usepackage{placeins}
\usepackage{bm}
\usepackage{amsmath}
\usepackage{amsfonts}
\usepackage{amssymb}
\usepackage[english]{babel}
\begin{document}

\preprint{AIP/123-QED}

\title{Thawed Gaussian Ehrenfest dynamics at conical intersections: When can a single mean-field trajectory capture internal conversion?}

\author{Alan Scheidegger}
\email{alan.scheidegger@epfl.ch}
\author{Ji\v{r}\'{\i} J. L. Van\'{\i}\v{c}ek}
\email{jiri.vanicek@epfl.ch}
\affiliation{Laboratory of Theoretical Physical Chemistry, Institut des Sciences et
Ing\'enierie Chimiques, Ecole Polytechnique F\'ed\'erale de Lausanne (EPFL),
CH-1015, Lausanne, Switzerland}
\date{\today}

\begin{abstract}
The thawed Gaussian Ehrenfest dynamics is a single-trajectory method that partially includes both nuclear quantum and electronically nonadiabatic effects by combining the thawed Gaussian wavepacket dynamics with Ehrenfest dynamics.
First, we demonstrate the improvement over the parent methods in a multidimensional system consisting of vertically displaced harmonic potentials with constant diabatic couplings, for which the thawed Gaussian Ehrenfest dynamics is exact.
Then, we show that single-trajectory mean-field methods completely fail to capture electronic population transfer in the vicinity of conical intersections between potential energy surfaces associated with electronic states of different symmetry (i.e., belonging to different irreducible representations of the molecular point group). The underlying cause of this limitation suggests that the thawed Gaussian Ehrenfest dynamics can be useful for studying nonadiabatic dynamics close to conical intersections of electronic states of the same symmetry, which have been understudied owing to the difficulty in locating them. Using a model of this type of intersection, we compare the thawed Gaussian Ehrenfest dynamics with exact quantum dynamics and find that the approximate mean-field approach yields a molecular wavefunction that remains qualitatively similar to the exact one even after crossing and recrossing the conical intersection.

\end{abstract}

\maketitle

\graphicspath{{./figures/}}

\section{Introduction}
The Born-Oppenheimer approximation simplifies molecular dynamics calculations by separating the treatment of nuclei and electrons, under the assumption that electrons adjust promptly to nuclear movements and remain in the same quantum eigenstate as nuclei evolve.
Within this framework, the nuclear wavefunction generally evolves on a single potential energy surface, but the approximation can be generalized to an initial superposition of electronic states, with each nuclear wavepacket evolving on its respective potential energy surface.
However, the Born-Oppenheimer approximation is inadequate for describing nonadiabatic dynamics, where interactions between electronic and nuclear motions become non-negligible.

Many methods have been developed to perform nonadiabatic dynamics while mitigating the exponential scaling of the exact grid-based quantum solution~\cite{Agostini_Curchod:2019}. Some of these, such as the multi-configurational time-dependent Hartree~\cite{Meyer_Cederbaum:1990,Beck_Meyer:2000} and variational multi-configurational Gaussian methods~\cite{Worth_Burghardt:2004}, multi-configurational Ehrenfest dynamics~\cite{Shalashilin:2009} and full multiple spawning~\cite{Martinez_Levine:1997}, remain exact in the limit of an infinite number of basis functions. More approximate mixed quantum-classical dynamics methods offer an efficient alternative by describing nuclear motion classically while treating electrons quantum-mechanically. In practice, some nuclear quantum effects can be recovered by propagating a swarm of trajectories and obtaining nuclear expectation values by an ensemble average. Mixed quantum-classical methods, including the multitrajectory Ehrenfest dynamics~\cite{Ehrenfest:1927,Billing:1983}, trajectory surface hopping~\cite{Tully:1990}, mapping approaches such as the Meyer--Miller model~\cite{Meyer_Miller:1979,Cotton_Miller:2017,Stock_Thoss:1997} and spin-mapping methods~\cite{Runeson_Richardson:2019,Mannouch_Richardson:2023,Mannouch_Kelly:2024}, and the coupled-trajectory mixed quantum-classical algorithm~\cite{Curchod_Tavernelli:2018} have been successfully applied to many molecular systems. In particular, they generally provide accurate short-time dynamics across conical intersections~\cite{Gherib_Izmaylov:2015}.

Nevertheless, all of these methods still suffer from an important computational overhead, as they require the propagation of multiple trajectories.
In this context, Ehrenfest dynamics in its single-trajectory formulation stands out as the most efficient, but also most approximate, mixed quantum-classical method. Single-trajectory Ehrenfest dynamics is limited by the overestimation of electronic coherence, as all electronic states share the same classical trajectory of the nuclei, and by the inability to account for nuclear quantum effects. In contrast, the single-trajectory semiclassical Gaussian wavepacket dynamics methods neglect nonadiabatic effects, but improve on classical dynamics by partially incorporating nuclear quantum effects through the nonzero width of the many-dimensional Gaussian representing the nuclear wavefunction~\cite{Heller:1975,Vanicek:2023}. These methods offer various levels of accuracy and computational cost. The simplest among them is the single-trajectory frozen Gaussian approximation, which associates a Gaussian wavepacket of a fixed width to the classical trajectory. Conversely, the variational thawed Gaussian approximation provides the optimal solution for a Gaussian wavepacket ansatz and, surprisingly, can qualitatively describe quantum tunneling~\cite{Moghaddasi_Vanicek:2023}.

The thawed Gaussian Ehrenfest dynamics (TGED) family of methods~\cite{Vanicek:2025} can be seen as a generalization of Ehrenfest dynamics and semiclassical Gaussian wavepacket dynamics, or alternatively, as an application of the time-dependent Hartree approximation~\cite{Dirac:1930} to various quadratic effective molecular potentials. A one-to-one correspondence links each TGED method to its single-surface thawed Gaussian wavepacket dynamics (TGWD) analog in the limit of uncoupled electronic states.

Here, we concentrate on the single-Hessian TGED, the nonadiabatic mean-field analog of the single-Hessian TGWD, where the Hessian of the effective potential along the nuclear trajectory is kept constant~\cite{Begusic_Vanicek:2019}.
We use the quadratic vibronic coupling Hamiltonian to parameterize the potential energy surfaces for different scenarios and evaluate the performance of TGED.
First, we present numerical results for a system in which the TGED is exact and highlight its advantages over both the TGWD and Ehrenfest dynamics. Next, we explain why single-trajectory mean-field methods completely fail to capture electronic population transfer near conical intersections between electronic states belonging to different irreducible representations of the molecular point group, using pyrazine as a case study. Importantly, most conical intersections studied in the literature fall into this category because their locations can be predicted using group theory.
Finally, we show that TGED is useful for studying nonadiabatic dynamics near conical intersections between electronic states of the same symmetry, which have been largely overlooked so far. In particular, conical intersections of asymmetric molecules (which belong to the $C_{1}$ point group) always fall into this category. We evaluate the performance of TGED in a system modeling such a conical intersection and demonstrate that the semiclassical mean-field trajectory provides a good description of the crossing and recrossing of the intersection.

\section{Theoretical background}
The systems used in this study are parameterized with the quadratic vibronic coupling Hamiltonian, which is often used to model potential energy surfaces of real molecules in the vicinity of conical intersections. In this section, we briefly review how this Hamiltonian is constructed and provide the equations of motion of TGED in the diabatic representation, which conveniently avoids the complexities associated with the geometric phase~\cite{Ryabinkin_Izmaylov:2014} and the divergence of derivative couplings at conical intersections.
\label{sec:theory}
\subsection{The quadratic vibronic coupling Hamiltonian}
The construction of the quadratic vibronic coupling Hamiltonian model~\cite{book_Domcke_Koppel:2004,Cattarius_Cederbaum:2001,Faraji_Koppel:2011} starts with a reference nuclear Hamiltonian
\begin{equation}
    \hat{H}_{0}=\hat{T}_{N}+\hat{V}_{0}=-\frac{1}{2}\sum_{i=1}^{D}\hbar \omega_{i}\frac{\partial^2}{\partial q_{i}^2}+\frac{1}{2}\sum_{i=1}^{D}\hbar\omega_{i}q_{i}^2,
\end{equation}
expressed as a sum of the nuclear kinetic energy $T_{N}$ and a quadratic expansion $V_{0}$ of the potential energy surface of a reference electronic state about a reference position taken to be at $q=(0, \ldots, 0)$. We shall consider the reference state to be the ground electronic state and the reference position to be the minimum of the electronic ground-state surface, which is also the location of the Franck-Condon point. Each dimensionless mass- and frequency-scaled coordinate $q_{i}$ is associated with mass $m_{i}$, force constant $k_{i}$, and frequency $\omega_{i}=\sqrt{k_{i}/m_{i}}$.
In the quadratic vibronic coupling model, the full molecular Hamiltonian describing a system with $S$ coupled electronic states takes the form
\begin{equation}
    \begin{split}
    \mathbf{\hat{H}}&=\hat{H}_{0}\mathbf{1}_{S}+\mathbf{W}(\hat{q}),
    \end{split}
\end{equation}
where $\mathbf{1}_{S}$ is the $S\times S$ unit matrix. The off-diagonal elements of $\mathbf{W}(q)$,
\begin{equation}
\label{eq:off_diagional_elements}
    W_{mn}(q) = W_{mn}^{0} + \sum_{i=1}^{D}\lambda_{imn}q_{i},\:\:\:\:\:\:\:m\neq n,
\end{equation}
linearly couple different electronic states $m$ and $n$, enabling transfer of electronic population.
Importantly, the constant, zeroth-order term $W_{mn}^{0}$ vanishes when the electronic states $m$ and $n$ transform according to different irreducible representations~\cite{Domcke_Koppel:2004}. The diagonal elements,
\begin{equation}
\label{eq:on_diagional_elements}
    \begin{split}
     W_{mm}(q) = E_{m} + \sum_{i=1}^{D}\kappa_{im}q_{i} + \sum_{ij}g_{ijm}q_{i}q_{j},
    \end{split}
\end{equation}
add three modifications to the harmonic reference potential $V_{0}(q)$ to define the $m$th electronic surface. The first term shifts the $m$th electronic surface vertically by $E_{m}$. The second term shifts the $m$th surface horizontally by $\Delta q_{i} = -\kappa_{im}/\hbar\omega_{i}$, which also leads to a nonzero gradient at the Franck-Condon point.
Finally, the last term, which is neglected in the linear vibronic coupling Hamiltonian~\cite{Penfold_Eng:2023}, induces changes in the vibrational frequencies (``mode distortion'') when $i=j$  and rotation of the normal coordinates (``Duschinsky effect''~\cite{Duschinsky:1937} or ``mode mixing'') when $i\neq j$.
The quadratic vibronic coupling Hamiltonian is commonly employed to simulate nonadiabatic dynamics involving conical intersections in polyatomic molecules.
When restricted to a single surface, the model reduces to the global harmonic approximation, where the TGWD and the TGED become equivalent and exact. This makes the quadratic vibronic coupling Hamiltonian a useful model for exploring the factors that determine the success or failure of the TGED method.

\subsection{Equations of motion of the TGED}
In the time-dependent Hartree approximation, the molecular wavefunction is written as the Hartree product
\begin{equation}
    \Psi(t)=\psi(t)\cdot \mathbf{c}_{t},
\end{equation}
where $\psi(t)$ represents the nuclear state and the $S$-dimensional complex vector $\mathbf{c}_{t}$ consists of the components of the electronic wavepacket in a chosen electronic basis.
The complex vector $\mathbf{c}_{t}$ also appears in Ehrenfest dynamics, where the nuclear part is represented by the position and momentum vectors $q_{t}$ and $p_{t}$, because the nuclear wavefunction is assumed to be infinitesimally narrow in phase space.
For this reason, the molecular wavefunction is no longer defined, and single-trajectory Ehrenfest dynamics does not capture nuclear quantum effects. In contrast, the TGED~\cite{Vanicek:2025} represents the nuclear state by a Gaussian wavepacket
\begin{align}
	\psi(q,t)&=\exp\bigg[
		\frac{i}{\hbar} \bigg(
		\frac{1}{2}x^{T} \cdot A_{t} \cdot x
		+ p_{t}^{T} \cdot x + \gamma_{t}
		\bigg)
	\bigg],\label{eq:gaussian_wavepacketP}\\
 x&:=q-q_{t},
 \end{align}
where $q_{t}$ and $p_{t}$ are the phase space coordinates of the center of the wavepacket, $A_{t}$ is a complex symmetric width matrix with a positive-definite imaginary part, and $\gamma_{t}$ is a complex number whose real part adds a dynamical phase and imaginary part ensures normalization at all times. The TGED is based~\cite{Vanicek:2025} on replacing the exact potential $\mathbf{V}$ with some locally quadratic effective potential $\mathbf{V}_{\mathrm{eff}}$. Here, we will use the single-Hessian TGED, where the approximate effective molecular potential is
\begin{equation}
\label{eq:molecular_potential}
\mathbf{V}_{\mathrm{eff}}(q;\psi)=\mathbf{V}_{0}(q_{t})+\mathbf{V}_{1}^{T}(q_{t})\cdot x+x^{T}\cdot\mathbf{V}_{2}(q_{r})\cdot x/2.
\end{equation}
Coefficients $\mathbf{V}_{0}$ and $\mathbf{V}_{1}$ are the diabatic potential energy matrix and its gradient at the center of the wavepacket $q_{t}$, while $\mathbf{V}_{2}$ is the Hessian of the potential energy matrix at a reference point $q_{r}$.
In the Hartree mean-field approach, the nuclear wavepacket~(\ref{eq:gaussian_wavepacketP}) is propagated with an effective nuclear potential
\begin{align}
\label{eq:eff_nu_pot}
    V_{n,\mathrm{eff}} &:= \mathbf{c}_{t}^{\dagger}\mathbf{V}_{\mathrm{eff}}\mathbf{c}_t = V_{n,0} + V_{n,1}^{T}\cdot x + x^{T}\cdot V_{n,2}\cdot\ x/2,\\
    V_{n,j} &= \mathbf{c}_{t}^{\dagger}\mathbf{V}_{j}\mathbf{c}_t,
\end{align}
obtained by averaging  the multistate potential~(\ref{eq:molecular_potential}) over the electronic wavepacket. Consequently, the time dependence of the nuclear Gaussian wavepacket is given by a set of differential equations~\cite{Vanicek:2025}
\begin{align}
    \dot{q}_{t}&=m^{-1}\cdot p_t,\\
    \dot{p}_{t}&=-V_{n,1},\\
    \dot{A}_{t}&=-A_{t}\cdot m^{-1}\cdot A_{t}-V_{n,2},\\
    \dot{\gamma}_{t}&=T(p_{t})-V_{n,0}+(i\hbar /2)\mathrm{Tr}(m^{-1}
    \cdot A_{t}),
\end{align}
identical to that of the TGWD, with the exception that the potential is $V_{n,\mathrm{eff}}$ instead of $V_{\mathrm{eff}}$, the single-surface analog of Eq.~(\ref{eq:molecular_potential}).
Conversely, the effective electronic potential matrix is obtained by averaging the effective molecular potential~(\ref{eq:molecular_potential}) over the nuclear wavepacket~(\ref{eq:gaussian_wavepacketP}):
\begin{equation}
\begin{split}
\label{eq:electronic_potential}
    \langle\mathbf{V}_{\mathrm{eff}}(q;\psi)\rangle_{n}
    &=\langle\psi|\mathbf{V}_{\mathrm{eff}}(q;\psi)|\psi\rangle\\
    &=\mathbf{V}_{0}(q_{t})+\mathrm{Tr}[\mathbf{V}_{2}(q_{r})\cdot \mathrm{Cov}(q)]/2.
\end{split}
\end{equation}
The position covariance matrix $\mathrm{Cov}(q)=(\hbar /2)(\mathrm{Im}A_{t})^{-1}$ adds a correction to the Ehrenfest electronic potential $\mathbf{V}_{0}$ due to the finite width of the nuclear wavepacket. In the TGED, the equation of motion for the electronic coefficients
\begin{equation}
    i\hbar\mathbf{\dot{c}}_{t}=[T(p_{t})+\langle\mathbf{V}_{\mathrm{eff}}(q;\psi)\rangle_{n}]\mathbf{c}_{t}
\end{equation}
resembles that of Ehrenfest dynamics, but the potential $\mathbf{V}(q_{t})$ is replaced by $\langle\mathbf{V}_{\mathrm{eff}}(q;\psi)\rangle_{n}$.

\section{Results and discussion}
\label{sec:results}
In this section, we begin by demonstrating the conditions under which TGED becomes exact and outperforms both Ehrenfest dynamics and the TGWD. Then, using pyrazine as a case study, we show that single-trajectory mean-field methods fail to capture electronic population transfer near conical intersections between electronic states of different symmetry and explain the underlying cause. Finally, we apply the TGED to a model representing a conical intersection between electronic states of the same symmetry. We show, and also explain why, here the TGED agrees qualitatively with exact quantum dynamics evaluated with the split-operator algorithm~\cite{Feit_Steiger:1982}.
\subsection{Vertically displaced harmonic oscillators}
\label{sec:HO}
The first example demonstrates when TGED is exact. Our model is loosely based on the system proposed by Tully and coworkers~\cite{Kohen_Tully:1998}, involving a heavy and a light particle, both represented by harmonic potentials, and mutually coupled with a bilinear coupling. As a result, the potential energy surfaces along the heavy particle coordinate correspond to equally spaced vertically displaced harmonic potentials.
Physically, this system topology can represent a harmonic vibration with electronic states broadened by the solvent~\cite{Fischer_Li:2011}.
In this model, Ehrenfest dynamics was demonstrated to produce more accurate results than surface hopping, which introduces an artificial loss of coherence~\cite{Fischer_Li:2011, Kohen_Tully:1998}.
However, one should remember that a mean-field trajectory generally underestimates electronic decoherence effects, since all electronic states share the same nuclear configuration. In contrast, when multiple ``locally'' mean-field trajectories are propagated~\cite{Zimmermann_Vanicek:2012}, although each individual path remains fully coherent, the total Hamiltonian no longer follows the mean-field approximation, enabling decoherence to occur.

Here, we consider two nuclear degrees of freedom and ten electronic states. Retaining the energy gaps of $\sqrt{5}$ a.u., vibrational frequency of $k_{i}=15$ a.u. and mass of $m_{i}=10$ of the original paper of Tully and coworkers, our system consists of ten vertically displaced two-dimensional harmonic surfaces centered at the origin. The diabatic couplings between each pair of electronic states are arbitrarily set to the constant value $W_{mn}=0.6$ a.u.
Because the diabatic potential energy surfaces are mutually parallel and because the diabatic couplings are constant functions of the nuclear coordinates, the electron-nuclear correlation vanishes, making this model particularly suited for mean-field methods using a single Hartree product to represent the molecular wavefunction.

\begin{figure}
\includegraphics[width=\columnwidth]{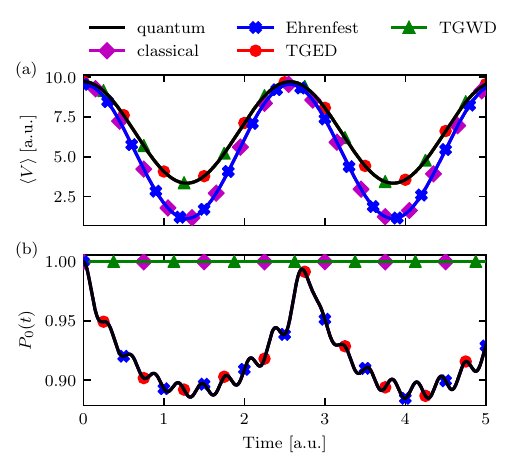}
\caption{(a) Time dependence of the expectation value $\langle V \rangle$ of the potential energy and (b) electronic ground-state population $P_{0}$. The figure compares the results of exact quantum dynamics, Ehrenfest dynamics, thawed Gaussian Ehrenfest dynamics (TGED), thawed Gaussian wavepacket dynamics (TGWD), and classical dynamics for a system of ten parallel two-dimensional harmonic potentials with constant diabatic couplings (see the main text for details).}
\label{fig:harmonic_S10}
\end{figure}
Figure \ref{fig:harmonic_S10} shows the time evolution of the expectation value of the potential energy and the time-dependent population on the ground electronic surface when the initial state, initially centered at $q_{0}=(-0.75, -0.75)$ with no momentum, is propagated with exact quantum dynamics, TGED, Ehrenfest dynamics, TGWD, or classical equations of motion.
The perfect overlaps of the black and red curves in panels (a) and (b) confirm that TGED is exact in vertically displaced harmonic potentials with constant diabatic couplings.

Surprisingly, Ehrenfest dynamics and classical dynamics yield the same expectation value for the potential energy [Fig.~\ref{fig:harmonic_S10}(a)]. This equivalence can be understood from the following observations:
(i) Since the diabatic surfaces are all parallel, with constant couplings between them, the effective nuclear potential of Eq.~(\ref{eq:eff_nu_pot}) is also parallel to the surface on which the classical particle evolves, resulting in the equivalence of the expectation value of the kinetic energies
\begin{equation}
\langle T_{\mathrm{Ehr}}(t)\rangle_{e} = T_{\mathrm{cl
}}(t),
\end{equation}
where $\langle \rangle_{e}$ denotes the integration over the electronic degrees of freedom.

(ii) Initially, the total energy is equal for classical and Ehrenfest dynamics. Additionally, both methods conserve the total energy $\langle H \rangle=\langle T \rangle+\langle V \rangle$, so we can write
\begin{equation}
H_{\mathrm{cl}}(t)=H_{\mathrm{cl
}}(0)=\langle H_{\mathrm{Ehr}}(0)\rangle_{e}=\langle H_{\mathrm{Ehr
}}(t)\rangle_{e}.
\end{equation}
(iii) Combining the first two observations straightforwardly leads to
\begin{equation}
\langle V_{\mathrm{Ehr}}(t)\rangle_{e}= V_{\mathrm{cl
}}(t).
\end{equation}
In other words, for Ehrenfest dynamics in this system, the contributions to the potential energy from electronic states inaccessible to the classical particle and the contributions due to the diabatic couplings cancel out.
The same reasoning can be applied to compare the TGED and the TGWD, both of which yield the exact expectation value of the potential energy [Fig.~\ref{fig:harmonic_S10}(a)].
Interestingly, in the adiabatic representation, the exact TGED corresponds to the simultaneous evolution of multiple nuclear wavepackets on several harmonic electronic surfaces while maintaining constant populations, as the nonadiabatic coupling is zero. This implies that the exact dynamics can also be reproduced by running one appropriately weighted TGWD trajectory per adiabatic state.

Of course, neither the TGWD nor classical dynamics can describe transfer of electronic population, which thus remains constant for both methods in panel (b). Despite the absence of a nuclear wavefunction, Ehrenfest dynamics yields the same population dynamics as the TGED (which is exact). To understand this, we need to see how the effective electronic matrix~(\ref{eq:electronic_potential}) of the TGED differs from that of Ehrenfest dynamics.
In this system, the second-order coefficients $g_{ijm}$ in Eq.~(\ref{eq:on_diagional_elements}) do not depends on $m$ (i.e., are independent of the electronic degree of freedom) and the diabatic couplings are constant [i.e., $\lambda_{imn}=0$ in Eq.~(\ref{eq:off_diagional_elements})]. Thus, the $S\times S\times N \times N$ second derivative tensor $\mathbf{V}_{2}$ in Eq.~(\ref{eq:molecular_potential}) can be expressed as a tensor product $\mathbf{V}_{2}=\mathbf{1}_S\otimes V_{2}$, where $V_{2}$ is the same $N \times N$ Hessian of each diabatic surface. The effective electronic matrix (\ref{eq:electronic_potential}) reduces to
\begin{equation}
\langle\mathbf{V}_{\mathrm{eff}}(q;\psi)\rangle_{n}=\mathbf{V}_{0}+\mathbf{1}_{S}\mathrm{Tr}[V_{2}\cdot \mathrm{Cov}(q)]/2.
\end{equation}
Consequently, the contribution from the finite width of the Gaussian wavepacket only shifts the diagonal elements of the electronic matrix $\mathbf{V}_{0}$ by the same amount.
As a result, the electronic coefficients evaluated with Ehrenfest dynamics or the TGED have different phases, but the electronic populations are identical.

We can conclude that the TGED is expected to provide accurate results when the electronic surfaces have similar shapes and the diabatic couplings have only a weak dependence on the nuclear coordinates. Ideally, these couplings should be constant, which can be an appropriate approximation in some cases~\cite{Guo:1993}.

\subsection{Internal conversion between electronic states of different symmetry}
\label{sec:MF_lim}
\begin{figure}
\includegraphics[width=\columnwidth]{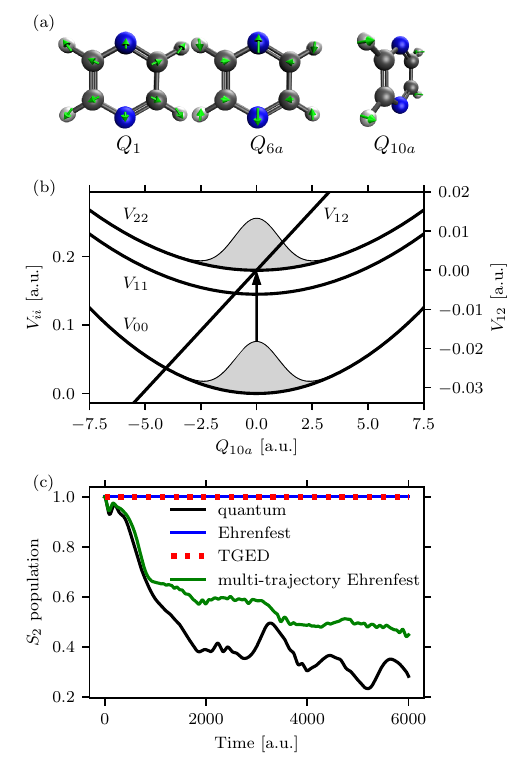}
\caption{(a) Vibrational normal modes $Q_{1}$, $Q_{6a}$, and $Q_{10a}$ of the electronic ground-state surface of pyrazine, (b) Cut of the diabatic potential energy surfaces and coupling along the coupling mode $Q_{10a}$. Excitation of the initial nuclear wavepacket is represented schematically. (c) Electronic population of the second diabatic excited state $S_{2}$, computed exactly, and compared with multi-trajectory Ehrenfest dynamics, single-trajectory Ehrenfest dynamics, and the thawed Gaussian Ehrenfest dynamics (TGED).}
\label{fig:pyrazine_pop_pot}
\end{figure}
Using pyrazine as an example, we now present a scenario where single-trajectory mean-field methods, including TGED, completely fail to capture electronic population transfer. Pyrazine belongs to the $D_{2h}$ symmetry group and contains a conical intersection between the potential energy surfaces of the first two excited adiabatic electronic states.
In the 3-dimensional, 3-state vibronic coupling Hamiltonian model constructed by Woywod \textit{et al.}~\cite{Woywod_Werner:1994}, the diabatic $S_{1}$ and $S_{2}$ electronic states transform according to the $B_{3u}$ and $B_{2u}$ irreducible representations, respectively, and are coupled through a vibrational mode $Q_{10a}$ of $B_{1g}$ symmetry. The two totally symmetric modes $Q_{1}$ and $Q_{6a}$, represented in Fig.~\ref{fig:pyrazine_pop_pot}(a) together with $Q_{10a}$, are also included in the model.

The cut of the potential, shown in Fig.~\ref{fig:pyrazine_pop_pot}(b), shows that the three potential energy curves share the same equilibrium position along $Q_{10a}$.
Combined with the linear diabatic couplings between the excited states, the symmetry-imposed geometry of the diabatic surfaces has significant consequences for single-trajectory mean-field methods, leading to a breakdown in their ability to properly describe nonadiabatic dynamics.
In particular, a single Ehrenfest trajectory initialized on $V_{22}$ at the Franck-Condon point feels no force in the direction of the $Q_{10a}$ mode and no diabatic coupling. Consequently, in the absence of initial momentum, population transfer cannot occur, as illustrated in Fig.~\ref{fig:pyrazine_pop_pot}(c).
Regrettably, the issue persists with TGED, despite the presence of a nuclear wavefunction. The reason is that the expectation value of the linear diabatic coupling in Eq.~(\ref{eq:electronic_potential}) vanishes for a Gaussian wavepacket centered at the Franck-Condon point. This illustrates that TGED only partially incorporates nuclear quantum effects. Conversely, multitrajectory Ehrenfest dynamics overcomes this limitation by initializing trajectories with positions and momenta randomly sampled from the initial Gaussian wavepacket, enabling population transfer. Compared with the exact result, multitrajectory Ehrenfest dynamics accurately captures the diabatic population of the $S_2$ state for the first 1000 a.u. and provides a reliable qualitative estimate up to 6000 a.u.

\subsection{Symmetry of conical intersections}
\label{sec:sym_CI}
The absence of nonadiabatic effects in pyrazine when single-trajectory mean-field methods are used is not an isolated exception, but occurs for the vast majority of conical intersections represented in the literature by the vibronic coupling Hamiltonian. Understanding this effect requires analyzing the selection rule that determines the existence of conical intersections and the impact on their geometry.

Conical intersections are named this way because they form a cone-like shape in the adiabatic representation within the two-dimensional $g{\text -}h$ branching plane~\cite{Yarkony:2001}, spanned by the energy difference gradient vector $g$ and the interstate coupling vector $h$. The corresponding nuclear coordinates are commonly referred to as the tuning mode $Q_{\mathrm{t}}$ and the coupling mode $Q_{\mathrm{c}}$. As explained by Yarkony~\cite{Yarkony:1998a}, the specific roles of these coordinates can be understood from a simple $2\times 2$ diabatic potential matrix,
\begin{equation}
    \mathbf{V}(Q_{\mathrm{t}},Q_{\mathrm{c}})=\begin{pmatrix}S(Q_{\mathrm{t}},Q_{\mathrm{c}})+G(Q_{\mathrm{t}}) & V(Q_{\mathrm{c}})\\V(Q_{\mathrm{c}}) & S(Q_{\mathrm{t}},Q_{\mathrm{c}})-G(Q_{\mathrm{t}})\end{pmatrix},
\end{equation}
representing a conical intersection. The eigenvalues of $\mathbf{V}$ are degenerate when
\begin{subequations}
\label{eq:G_V_zero}
\begin{align}
    G(Q_{\mathrm{t}}) &= 0 \label{eq:G_zero},\\
    V(Q_{\mathrm{c}}) &=0 \label{eq:V_zero},
\end{align}
\end{subequations}
i.e., at a position where the diabatic coupling is zero along $Q_{\mathrm{c}}$ and where the diabatic curves cross along $Q_{\mathrm{t}}$.

Depending on the role of symmetry in solving Eqs.~(\ref{eq:G_zero}) and ~(\ref{eq:V_zero}), conical intersections are classified into three categories~\cite{Yarkony:1998a}: symmetry-required, symmetry-allowed, and same-symmetry conical intersections. The first category pertains to molecules with a non-Abelian point group, in situations where the electronic states defining the intersection belong to a degenerate irreducible representation. In this case, both conditions~(\ref{eq:G_zero}) and~(\ref{eq:V_zero}) can be determined from group theory. For symmetry-allowed intersections, only~(\ref{eq:V_zero}) can be found using symmetry. Finally, for same-symmetry intersections, both conditions are fulfilled accidentally~\cite{Yarkony:1996}. This explains why conical intersections between electronic states of the same symmetry have not been extensively studied, despite the development of algorithms to find them~\cite{Manaa_Yarkony:1990,Manaa_Yarkony:1993,Maeda_Morokuma:2009,Maeda_Morokuma:2010}.
In contrast, many diabatic models for conical intersections between electronic states of different symmetry can be found in the literature~\cite{Woywod_Werner:1994,Faraji_Koppel:2011, Baldea_Koppel:2006}.

Let us return to pyrazine to better understand how point group theory can be used to locate a symmetry-allowed conical intersection by solving Eq.~(\ref{eq:V_zero}). For all types of intersection, a nuclear normal mode $j$ can couple two electronic states $m$ and $n$ at first order if the relation~\cite{Worth_Cederbaum:2004,Neville_Schuurman:2022}
\begin{equation}\label{eq:selection_rule}
    \Gamma_{m} \otimes \Gamma_{j} \otimes \Gamma_{n} \supset\Gamma_{\mathrm{TS}},
\end{equation}
is satisfied. That is, if the direct product of the irreducible representations $\Gamma_{m}$ and $\Gamma_{n}$ of the two electronic states  and of the irreducible representation $\Gamma_{j}$ of the nuclear normal mode contains the totally symmetric irreducible representation $\Gamma_{\mathrm{TS}}$. In pyrazine, the only vibrational mode that can couple the electronic states $S_{1}$ $(B_{3u})$ and $S_{2}$ $(B_{2u})$ is the out-of-plane mode $Q_{10a}$ of $B_{1g}$ symmetry~\cite{Woywod_Werner:1994}. In addition, we saw that the diabatic coupling~(\ref{eq:off_diagional_elements}) between electronic states of different irreducible representations is a linear (and not an affine) function of the coupling mode. Consequently, the conical intersection must be contained in the $Q_{10a}=0$ hyperplane to satisfy Eq.~(\ref{eq:V_zero}). In contrast, the diabatic coupling between electronic states of the same symmetry evaluated to first order in $Q_{\mathrm{c}}$ is, in general, an affine function of the coupling coordinate (i.e., contains not only a linear but also a constant term), and the point at which the coupling vanishes cannot be determined from group theory.

Most conical intersections studied in the literature involve internal conversion between electronic states belonging to different irreducible representations, for which the coupling vibrational mode cannot be totally symmetric. This means that motion along this mode breaks the molecular symmetry. In Abelian point groups such as $D_{2h}$, positive and negative displacements along a non-totally symmetric mode result in the same configuration~\cite{Vester_Kuleff:2023}.
As a result, the potential energy curves along the coupling mode $Q_{\mathrm{c}}$ are symmetric with respect to the $Q_{\mathrm{c}}=0$ hyperplane, as seen for pyrazine in Fig.~\ref{fig:pyrazine_pop_pot}(c), and the intra-state coupling coefficient $\kappa_{im}$ in Eq.~(\ref{eq:on_diagional_elements}) of the coupling mode must be zero.
In contrast, the coupling mode between electronic states of the same symmetry must be totally symmetric, allowing the corresponding coefficient $\kappa_{im}$ to be nonzero.

\subsection{Internal conversion between electronic states of the same symmetry}
%
%
\begin{figure}
\includegraphics[width=\columnwidth]{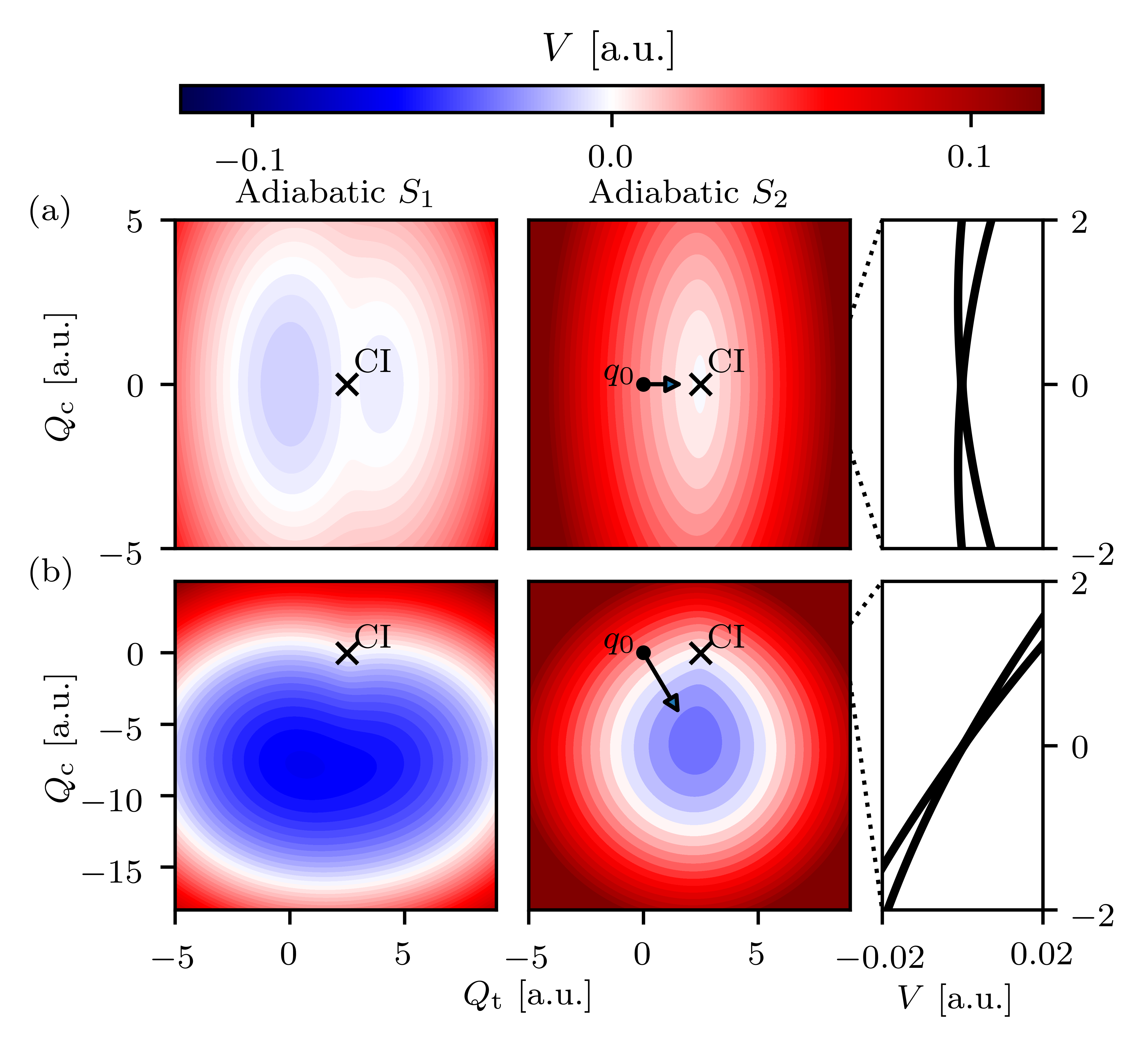}
\caption{Contour plots of the adiabatic surfaces defining conical intersections between electronic states of (a) different and (b) same symmetry. The cross marks the location of the intersection, and the arrow shows the direction of the force acting on a classical particle initialized at the Franck-Condon point $q_{0}$ in the excited state $S_{2}$. The last column presents cuts of the conical intersections along the coupling mode, highlighting their (a) peaked profile and (b) sloped profile.}
\label{fig:peaker_sloped_CI}
\end{figure}
\begin{figure*}
\includegraphics[width=\textwidth]{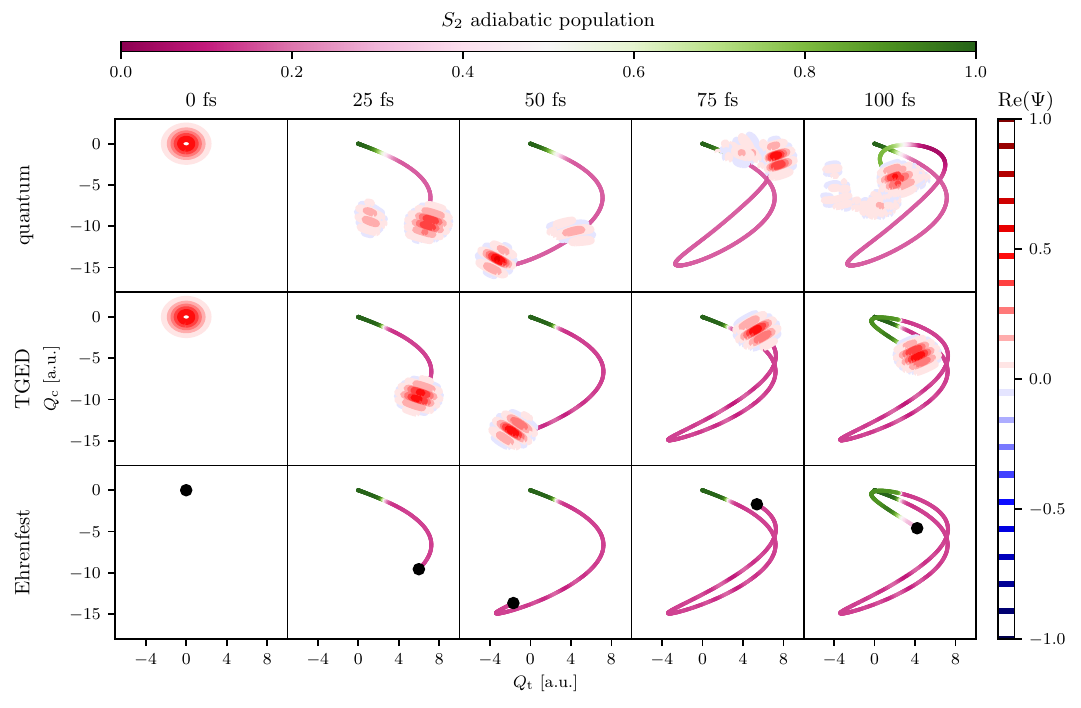}
\caption{Comparison of exact quantum dynamics, thawed Gaussian Ehrenfest dynamics (TGED), and Ehrenfest dynamics near a conical intersection of electronic states of the same symmetry. The real part of the nuclear wavepackets is shown in shades of blue when negative and red when positive. The adiabatic population of the $S_{2}$ electronic state is color-coded on the trajectory of the position expectation value.}
\label{fig:CI_mol_wf}
\end{figure*}
In Ref.~\onlinecite{Neville_Schuurman:2022}, Neville \textit{et al.} introduced two-dimensional models to represent conical intersections between electronic states of both different and same symmetry. The authors achieved this change of symmetry by varying the coefficient $\kappa_{im}$ of the coupling mode. Increasing this coefficient tilts the angle $n_{c}$ of the conical intersection. Using their parametrization, we took the models corresponding to peaked ($n_{c}=0^{\circ}$) and sloped ($n_{c}=10^{\circ}$) conical intersections. Although the diabatic potential is convenient for propagation, interpreting the dynamics based on the topology of the potential energy surfaces is not intuitive in this representation. For this reason, Fig.~\ref{fig:peaker_sloped_CI} shows the corresponding adiabatic potential surfaces defining the conical intersections, which are marked by a cross. Assuming an initial photoexcitation from the equilibrium position $q_{0}$ to the excited state $S_{2}$, the peaked and sloped conical intersections are expected to induce significantly different dynamics. In the first case, the nuclei are pushed directly in the direction of the intersection, which is why peaked conical intersections tend to facilitate population transfer between adiabatic states~\cite{Malhado_Hynes:2016}. However, a classical trajectory would remain on the straight line $Q_{\mathrm{c}}=0$, where the diabatic coupling is zero, thus preventing population transfer to other electronic states when using single-trajectory mean-field methods, as observed in the case of pyrazine in Sec.~\ref{sec:MF_lim}. In contrast, for electronic states of the same symmetry, the force on the nuclei can have nonzero components along both nuclear coordinates $Q_{\mathrm{t}}$ and $Q_{\mathrm{c}}$. Consequently, the nuclei do not pass directly through, but close to the conical intersection, and activation of the $Q_{\mathrm{c}}$ mode can induce electronic transitions.

It is important to note that the failure of Ehrenfest dynamics is not just an artifact of the diabatic representation. The adiabatic Hamiltonian can be expressed as
\begin{equation}
\label{eq:adiab_H}
    \mathbf{\hat{H}}_{\mathrm{ad}}=\frac{1}{2M}[\hat{p}^{2}\mathbf{1}-2i\hbar\mathbf{F}_{\mathrm{ad}}(\hat{q})\cdot \hat{p}-\hbar^{2}\mathbf{G}_{\mathrm{ad}}(\hat{q})]+\mathbf{V}_{\mathrm{ad}}(\hat{q})
\end{equation}
where $\mathbf{F}_{\mathrm{ad}}$ is an $S\times S$ matrix of nonadiabatic coupling vectors, $\mathbf{V}_{\mathrm{ad}}$ is the diagonal adiabatic potential energy matrix and $\mathbf{G}_{\mathrm{ad}}$ is an $S\times S$ matrix of scalar nonadiabatic couplings [some researchers only take the diagonal part, called the diagonal Born-Oppenheimer correction (DBOC)~\cite{Gherib_Izmaylov:2016}].
Even when the off-diagonal components of $\mathbf{G}_{\mathrm{ad}}$ are neglected, the term $\mathbf{F}_{\mathrm{ad}}(\hat{q})\cdot \hat{p}$ couples the adiabatic electronic states, enabling population transfer between them.
Figure~\ref{fig:peaker_sloped_CI}(a) indicates that this scalar product is initially zero for single-trajectory mean-field methods in the case of a conical intersection between electronic states of different symmetry, as $[\mathbf{F}_{\mathrm{ad}}]_{12}$, which is parallel to the $Q_{\mathrm{c}}=0$ axis, is orthogonal to the momentum $p$ (which is parallel to the $Q_{\mathrm{t}}=0$ axis). Approaching the conical intersection, the nonzero component of $[\mathbf{F}_{\mathrm{ad}}]_{12}$ diverges but the scalar product with $p$ remains zero. However, at the intersection itself, the component of $[\mathbf{F}_{\mathrm{ad}}]_{12}$ along $Q_{\mathrm{t}}$ takes the indeterminate form $\frac{0}{0}$, which could lead to numerical instability~\cite{Choi_Vanicek:2019}.
In contrast, when multiple trajectories are used, the randomly generated momentum vectors are generally not orthogonal to $[\mathbf{F}_{\mathrm{ad}}]_{12}$, enabling electronic transitions between the adiabatic states.

After observing the failure of single-trajectory mean-field methods to treat nonadiabatic dynamics between electronic states of different symmetry, we assess here the performance of Ehrenfest dynamics and the TGED in the model represented in Fig.~\ref{fig:peaker_sloped_CI}(b). This model describes a conical intersection between electronic states of the same symmetry, or more precisely, belonging to the same irreducible representation of the molecule's point group. In this case, the gradient at the Franck-Condon point has nonzero components along both nuclear coordinates. As a result, the momentum vector acquires a nonzero component along $Q_{\mathrm{c}}$, enabling population transfer even when using single-trajectory mean-field methods.
Figure \ref{fig:CI_mol_wf} shows the nonadiabatic dynamics computed exactly, with the TGED, and with Ehrenfest dynamics. The evolution of the position expectation value is displayed and color-coded to indicate the adiabatic population of the upper surface. A first population transfer to the lower surface occurs within the first 25 fs, as the nuclei pass near the conical intersection. 
The mean-field methods are in good agreement with the exact result. However, since the population transfer is not complete, a small portion of the exact wavepacket remains on the excited-state surface.
Although this splitting of the electronic wavepacket can be captured by both TGED and Ehrenfest dynamics, neither mean-field method can describe the subsequent splitting of the nuclear density due to the evolution on two different surfaces. Nevertheless, the nuclear wavepacket propagated with TGED remains in a qualitative agreement with the exact result up to 75 fs. After this time, a second crossing of the conical intersection occurs causing repopulation of the $S_{2}$ electronic state. Once again, the exact result reveals a splitting that leads to a more intricate structure of the nuclear wavepacket. Despite this complexity, the Gaussian of the TGED method resembles the exact nuclear wavefunction both in density and phase, which are absent in Ehrenfest dynamics.

\begin{figure}
\includegraphics[width=\columnwidth]{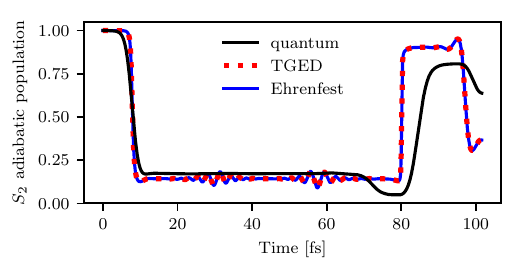}
\caption{Electronic population of the second excited adiabatic state $S_{2}$, computed exactly and compared with the thawed Gaussian Ehrenfest dynamics (TGED) and Ehrenfest dynamics.}
\label{fig:peaker_sloped_CI_population}
\end{figure}
Finally, Fig.~\ref{fig:peaker_sloped_CI_population} provides a more detailed view of the evolution of the electronic population of the $S_{2}$ adiabatic state.
First, we observe that Ehrenfest dynamics and the TGED yield identical population dynamics, for the same reason discussed in Sec.~\ref{sec:HO}. Overall, the single-trajectory mean-field methods agree well with the exact result. Nevertheless, the population transfers at 10 and 80 fs are significantly more abrupt compared to the exact reference.
Another notable difference is the appearance around 30 and 60 fs of rapid small-amplitude oscillations of the population evaluated with the TGED and Ehrenfest dynamics. This well-known effect is caused by the overestimation of electronic coherence caused by the mean-field approximation~\cite{Akimov_Prezhdo:2014,Ma_Burghardt:2018}. After the second crossing of the conical intersection, the TGED and Ehrenfest dynamics significantly diverge from the exact result.

\section{Conclusions and outlook}
\label{sec:conclusions}
The TGED improves on single-trajectory Ehrenfest dynamics by describing the nuclear part with a Gaussian wavepacket instead of a classical particle~\cite{Vanicek:2025}. Consequently, the total molecular wavefunction is fully defined, and both nonadiabatic effects and nuclear quantum effects are taken into account, at least partially.

After demonstrating the exactness of the TGED in vertically displaced harmonic potentials with constant diabatic couplings, we assessed the accuracy of the method in systems modeling different types of conical intersections. We have shown that mean-field methods fail to describe population dynamics in conical intersections between electronic states of different symmetry when the molecular point group is Abelian.
This behavior arises from the requirement that the coupling mode must be non-totally symmetric, which imposes that the energy gradients along the coupling mode are zero at the Franck-Condon point for all electronic surfaces. In contrast, when the electronic states have the same symmetry, the coupling mode is totally symmetric, allowing for a nonzero gradient of the potential energy. In this case, we show that the molecular wavefunction propagated with TGED can provide a qualitative picture of the exact dynamics over a duration that includes two crossings of the conical intersection.

Conical intersections between electronic states belonging to the same irreducible representation have received limited attention because of the challenges in locating them. Notably, this includes all intersections in molecules without any particular symmetry.
We propose using the TGED to efficiently capture the initial nonadiabatic dynamics near such intersections.
Additionally, the TGED method can be applied in cases where single-trajectory Ehrenfest dynamics is sufficient, with the potential for increased accuracy by incorporating some nuclear quantum effects.

\bibliographystyle{apsrev4-2}
\bibliography{biblio62,addition_application_TGED}

\end{document}